\documentclass[sigconf]{acmart}

\usepackage[utf8]{inputenc}

\usepackage{natbib}
\usepackage{graphicx}   
\usepackage{enumerate}
\usepackage{multirow}
\usepackage{booktabs}
\usepackage{todonotes}

\begin{document}

\title{Stateful Detection of Black-Box Adversarial Attacks}

\author{Steven Chen}
\affiliation{%
  \institution{University of California, Berkeley}
}

\author{Nicholas Carlini}
\affiliation{%
  \institution{Google Research}
}

\author{David Wagner}
\affiliation{%
  \institution{University of California, Berkeley}
}

\begin{abstract}
    The problem of \emph{adversarial examples}, evasion attacks on machine learning classifiers,
    has proven extremely difficult to solve.
    This is true even in the black-box threat model, as is the case 
    in many practical settings. 
    Here, the classifier is hosted as a remote service and and
    the adversary does not have direct access to the model parameters.
    %
    
    This paper argues that in such settings, defenders have a much larger space
    of actions than have been previously explored.
    Specifically, we deviate from the implicit assumption made by prior work that a
    defense must be a stateless function that operates on individual examples,
    and explore the possibility for stateful defenses.
    
    To begin, we develop a defense designed to detect the process of
    \emph{generating} adversarial examples.
    By keeping a history of the past queries, a defender can try to identify when a sequence of queries
    appears to be for the purpose of generating an adversarial example.
    We then introduce \emph{query blinding}, a new class of attacks designed to bypass defenses that
    rely on such a defense approach.
    
    We believe that expanding the study of adversarial examples from 
    stateless classifiers to stateful systems is not only more realistic
    for many black-box settings, but also
    gives the defender a much-needed advantage in responding to the adversary.
    
\end{abstract}

\maketitle
\vspace{1em}

\section{Introduction}

Over the past few years, neural networks have driven advancement in a wide range of domains. Deep learning based methods have achieved state of the art performance in areas including, gameplaying AIs for Go and chess \cite{AlphaGo}, machine translation between different languages \cite{MachineTranslation}, and classification and object detection for images \cite{ObjDetection}. 

Accordingly, neural networks are also increasingly used in safety-critical applications, where the reliable performance of these networks and their security against a malicious adversary is paramount. 
In some cases, such as a local image recognition system, the network and its parameters may be available to the adversary (the white-box case). 
However, when the classifier is hosted remotely (e.g., as a cloud service), only the output of the neural network is available to the adversary (the black-box case).

Worryingly, these neural networks have been shown to be highly vulnerable to \emph{adversarial examples}: inputs crafted by an adversary to deliberately fool a machine learning classifier.
Defending against adversarial examples has proven to be extremely difficult.
Most defenses that have been published have been found to have significant flaws \cite{carlini2017adversarial,obfuscated-gradients}, and even those few defenses that have thus far withstood validation offer only partial robustness \cite{madry2017towards}.

Adversarial examples might not actually be problematic in practice, where models are often held private by companies and hosted on the cloud.
However, surprisingly, adversarial examples can even be generated in a fully \emph{black-box} threat model.
Such an adversary in this threat model can only make queries of the model and receive the predicted classification label as output.
While there certainly are domains where the adversary will have white-box access to a deployed neural network, in many production environments when neural networks are deployed, the user is only allowed to make queries of the classifier and observe the output.
For example, services such as Clarifai \cite{Clarifai} and Google Cloud Vision AI \cite{GoogleVision} offer image classification APIs where users can submit images and receive only the label of that image. Similarly, for spam classification, a feature offered by many email providers, a user cannot directly access the actual spam classifier, but only observe whether an email is classified as spam or not.

This paper studies the problem of detecting the \emph{generation} of adversarial examples,
as opposed to trying to (statelessly) detect whether or not any individual input is malicious---which has proven to be difficult \cite{carlini2017adversarial}.
To do this, we consider the task of identifying the
\emph{sequence} of queries made to the classifier when creating an adversarial example.
The central hypothesis we evaluate is whether the sequence of queries used to generate a black-box adversarial example is distinguishable from the sequence of queries when under benign use.

Operating under this hypothesis, we propose a defense that relies on the specific observation that existing black-box attacks often make a 
sequence of highly self-similar queries (i.e., each query in the sequence is similar to prior queries in the sequence).
We train a similarity-detector neural network to identify such query patterns,
and find that the existing state-of-the-art black-box adversarial example attack algorithms can be easily detected through this strategy.
Our proposed strategy can trivially compose with any existing defense for defense-in-depth.

Then, we study adaptive attacks against our scheme, to understand whether an attacker who is aware of our detection strategy
could evade it.
We develop \emph{query blinding}, a general strategy for attacking defenses which monitor the sequence of queries in order to detect adversarial example generation.
Query blinding attacks pre-process each input with a \emph{blinding function} before querying the classifier, so that (1) the pre-processed inputs match the benign data patterns, but (2) it is possible to deduce the classifier's output from the result of these queries.
We show that our defense remains secure against query blinding.

Given the difficulty in defending or detecting attacks statelessly,
we believe that this new research direction---stateful methods for detecting black-box attacks---presents renewed hope for defending against adversarial example attacks in the black-box threat model.


We make the following contributions:
\begin{itemize}
    \item We propose a new class of adversarial example defenses: stateful detection defenses.
    
    \item We design and evaluate a defense in this category, and find it is
    effective at detecting existing attacks and is hard to evade even when the
    attacker adapts the attack approach.
    
    \item We introduce \emph{query blinding}, a general strategy that can be used
    to attack stateful detection defenses.
    
    \item We release the source-code for our defense and attacks at
    \textbf{https://github.com/schoyc/blackbox-detection}.
    
\end{itemize}




\section{Background \& Problem Statement}
This paper studies evasion attacks on neural networks \cite{Biggio}, commonly referred to as adversarial examples \cite{Szegedy}.
\subsection{Preliminaries} \label{subsection:preliminaries}

\bigskip \noindent
\textbf{Neural Networks.} 
A neural network is a function $f(\cdot)$ consisting of multiple layers. Each layer computes a weighted linear combination of the outputs of the previous layer, followed by a non-linearity. Because neural networks can have arbitrarily many layers, as well as various non-lineariaties, they can provably approximate arbitrarily complicated functions.
The \emph{weights} $\theta$ of a neural network refer to the parameters used in the weighted linear combinations.
To be explicit we may write $f_\theta(\cdot)$ to refer to the network $f(\cdot)$ with weights $\theta$.
Most of the recent impressive results in machine learning have come from applying neural networks \cite{AlphaGo,MachineTranslation,ObjDetection}.
This paper focuses on \emph{classification} neural networks, where some example $x$ is processed by the neural network to return the predicted \emph{label} $y=f(x)$ of the example.

\bigskip \noindent
\textbf{Training Neural Networks.}
A neural network begins with randomly initialized weights $\theta$.
Training a neural network is the process of iteratively updating the
weights to better solve the given task.

Neural networks are most often trained with stochastic gradient descent.
Given a set of labeled training examples $\mathcal{X}$ with examples $x_i$ and corresponding labels $y_i$,
gradient descent attempts to solve
\[\mathop{\text{arg min}}_\theta \mathop{\mathbb{E}}_{(x,y) \in \mathcal{X}} \ell\big(f_\theta(x), y\big)\]
where $\ell(\cdot)$ measures the \emph{loss}: intuitively, how ``wrong'' the prediction $f_\theta(x)$ is compared to the true label $y$.
Stochastic gradient descent (SGD) solves the above problem by iteratively updating the weights
\[\theta \leftarrow \theta - \varepsilon \cdot \nabla_\theta\bigg(\sum_{(x,y) \in \mathbb{B}} \ell\big(f_\theta(x),y\big)\bigg) \]
where $\nabla_\theta$ is the gradient of the loss with respect to the weights $\theta$; $\mathbb{B} \subset \mathcal{X}$ is a randomly selected \emph{mini-batch} of training examples drawn i.i.d. from $\mathcal{X}$; and $\varepsilon$ is the \emph{learning rate} which controls by how much the weights $\theta$ should be changed.

\bigskip \noindent
\textbf{Adversarial Examples.}
To formalize the definition of adversarial examples, most existing work \cite{AdvExamples,madry2017towards,obfuscated-gradients} defines an adversarial example as an input $x'$, which is a slightly modified version of a naturally occurring example $x$, such that a neural network classifies them differently.
Formally, an adversarial example $x'$ satisfies two properties: (1) for some $d(\cdot)$, a distance metric, $d(x, x') < \varepsilon$, but (2) for the neural network, $f(x) \neq f(x')$.
As long as $\varepsilon$ is set to be small enough, the perturbation that is introduced should not change the actual true classification of the object in the image (e.g. a dog with small perturbations is still a dog).

\bigskip \noindent
\textbf{Generating Adversarial Examples.}
The problem of generating adversarial examples can be formalized as a minimization problem
\[\delta^* = \mathop{\text{arg max}}_\delta \ell(f(x+\delta), y)\]
subject to the constraint that $\lVert\delta\rVert$ is small according to some metric.

White-box attacks to generate adversarial examples largely rely on using the same gradient-descent process used to train a neural network. Initially, we set $\delta_0=0$ and then update 
\[ \delta_{i+1} = \delta_i + \nabla_x \ell(f(x+\delta_i),y) \]
for some chosen loss function $\ell$.

Black-box attacks, by definition, are unable to perform the above optimization because they are not able to compute the gradient of the loss.
Instead, black-box attacks must perform gradient-free optimization. 
This paper considers two possible state-of-the-art black-box attacks: NES \cite{NES} and the BoundaryAttack \cite{BoundaryAttack}.
While their implementations differ (significantly so), at a very high level they both rely on the same strategy.
Starting from some initial perturbation $\delta_0$, the attacks slowly query the classifier on a sequence of perturbations $f(x+\delta_i)$, each highly similar to the previous, with the objective of finding an input that is (a) misclassified and (b) introduces a small distortion. 

A thorough understanding of these black-box attacks is not necessary yet; we defer a complete description to Section \ref{section:attacks-description}.

\bigskip \noindent
\textbf{Problem Domain: Image Classification.}
Following most prior work on the space of adversarial examples \cite{obfuscated-gradients}, this paper studies the domain of image classification.
Here, images are represented as $h\cdot{}w\cdot{}c$ dimensional vectors (with height $h$, width $w$, and $c$ color channels) drawn from $[0,1]^{hwc}$.

\subsection{Threat Model}
As discussed earlier, we focus on detecting black-box attacks for crafting adversarial examples.
In a black-box threat model, the adversary can query the model on any input and learn its classification, but the weights and parameters of the model are not released.
We envision, for example, a platform that makes available a machine learning model as a service, where the model can be queried by a user after he/she creates an account, but cannot download the model itself.
Under this threat model, we aim to increase the difficulty for attackers to craft adversarial examples.
While an attacker can query the model any number of times in trying to generate an adversarial example,
our goal is to detect such attacks before they are successful.

We focus on an account-oriented setting, where users must create an account before they can query the model.
Attackers might be able to create as many accounts as they wish, but we assume there is some practical cost associated with creating each account (e.g., linking to a valid credit card or phone number, paying an account fee, etc.).
In our scheme, the attacker's account is cancelled as soon as an attack-in-progress is detected, requiring the attacker to create a new account at that point.
A key metric for the effectiveness of our defense is the number of accounts that an attacker must create to successfully craft an adversarial example.
Each time the attack is detected, the attacker must create a new account, so we measure this by counting the number of times the attack is detected before it is successful (number of detections); this determines the attacker's cost to defeat the system.

Notice that our goal of detecting when an adversarial attack is in progress over a sequence of queries to the model is different from detecting whether or not any individual input is adversarial (as in previous detection based defenses \cite{carlini2017adversarial}).
Thus, our scheme involves retaining history of prior queries and scanning this history to check for patterns that indicate if an attack is in progress.
Such a defense is not feasible in the white-box threat model, where the defenders have no visibility into the attacker's offline computation.

We focus on the \emph{hard-label} setting, where each query to the model returns only the categorical label assigned by the classifier, but not a numerical confidence score associated with it.
Our approach would extend naturally to other settings, but as argued in prior work \cite{Biggio} we believe the hard-label setting is the most realistic black-box threat model.
(We have some evidence (see Appendix~E) that solving the soft-label setting may be more challenging.)

There are two broad types of black-box attacks in the literature: \emph{query-based attacks}, which make a sequence of queries to the model, and \emph{zero-query attacks}, which work entirely offline without interacting with the model.
While significant prior work has been dedicated to constructing defenses against the latter \cite{EAT}, limited work studies defenses against query-based attacks.
Our main contributions lie in studying defenses against query-based attacks. Our approach, monitoring the sequence of queries against the classifier, by definition can not detect zero-query attacks.

\subsection{Query-Based Attacks}
Our defense is motivated by the sequential nature of query-based black box adversarial attacks, such as NES \cite{NES} and the Boundary Attack \cite{BoundaryAttack}.
Query attacks iteratively perturb a source example to slowly transform it into an adversarial example according to some policy, usually by estimating gradients or boundary proximity.
This information is inferred by querying points near the current proposed adversarial example.

Considering attack queries as a sequence, successive queries are likely to be close together (by some distance metric), because (1) each iteration of the attack makes a small gradient-estimation step or boundary-following step from the current proposed adversarial example to the next proposed example; and (2) since only labels are accessible, the attack requires querying a random sample of points near the current example to approximate the actual gradient or decision boundaries of the model. Therefore, a scheme that tracks the sequence of queries made to a model can detect an attack based on an anomalous pattern of suspiciously close queries. 

As a concrete example, to generate an adversarial example of an input $x$, 
the label-only version of the NES attack \cite{NES} seeds
the attack with an input that is a different class than $x$, and then sequentially takes ``gradient'' steps from the adversarial image towards the original image to
(1) reduce the distortion from the current example to the original example $x$,
but (2) so that the class remains the target class.
Because obtaining the true gradient is not possible in the hard-label setting, the attack uses NES, a gradient-free optimization method, which estimates the gradient using the softmax scores of a random sample of points nearby the original example $x$.
For this gradient estimation to be accurate, queries must necessarily be made within a small distance of each other, which creates a pattern that we will use to detect the NES attack.

\subsection{Zero-Query Attacks}
One of the most surprising properties of adversarial examples is their \emph{transferability} \cite{AdvExamples}: given two different models (even trained on different datasets) for the same task, it turns out that adversarial examples generated on one will often transfer to the other.
This observation motivated the earliest black-box attack algorithms: train a ``surrogate model'' \cite{papernot2016transfer} on the same task as the target model, perform a gradient-based attack on the surrogate model, and replay this generated adversarial example on the target model.
This zero-query attack, while not $100\%$ successful, is surprisingly effective.

As mentioned earlier, our approach cannot defend against transfer attacks and other zero-query attacks.
Fortunately, others have proposed possible defenses to transfer attacks.
Perhaps the best known example is Ensemble Adversarial Training (EAT) \cite{EAT} which has been shown to be effective against zero-shot adversarial attacks.

The major limitation of zero-query defenses (including EAT) is that they are not effective against query attacks.
Thus, this prior work of defenses targeting the zero-query threat model perfectly complements our approach: we envision combining
our defense (to detect query-based attacks) with an existing defense (to detect zero-query attacks).
In Section \ref{section:zero-query-defense} we combine EAT with our defense to develop a complete defense to black-box adversarial examples.

\section{Our Scheme}
We now introduce and explain our scheme to detect black box, query based, adversarial attacks by tracking the sequence of queries the attacker makes in the process of generating an adversarial example.

\subsection{The Query Detection Defense}

\begin{figure*}[h!]
  \begin{center}
  \includegraphics[width=0.8\textwidth]{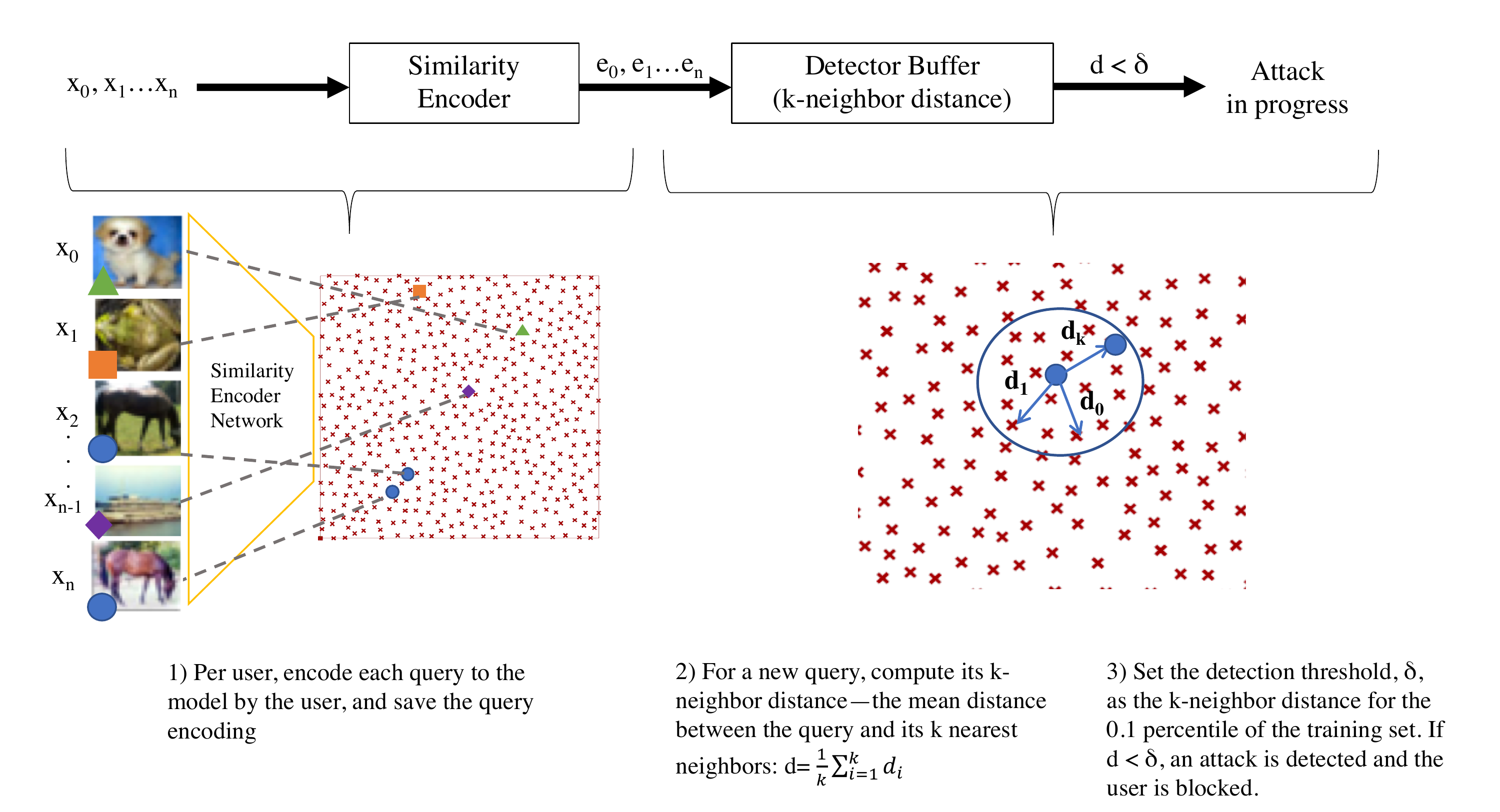}
  \caption{\textbf{Query Detection Defense:} The high-level process for detecting a query-based adversarial attack.}
  \label{fig:scheme}
  \end{center}
\end{figure*}

At a high level, our defense is applied as an access monitor on top of an existing classifier.
The detector records all queries to the classifier and stores them in a temporary history buffer.
For each new query, the detector computes the number of ``nearby'' examples in this temporary history buffer.
If we determine there are too many nearby examples, we report this as part of an attack sequence and take appropriate action (e.g., block this user's account).

In more detail,
for each user, we save every query from that user for a bounded duration (this period can be tuned according to the defender's resources, for example either a fixed amount of time or a fixed number of queries). 
Then, for each new query the system receives, we compute its \emph{k-nearest-neighbor distance} to the previously seen examples---the mean pairwise distance between the query and its $k$ nearest neighbors among the previously saved queries
(i.e., for each of the $k$ nearest neighbors, we compute the distance between the neighbor and query, then take the mean over these $k$ distances).

To measure the distance between queries, we encode the queries using a deep similarity encoder \cite{bell15productnet} (that maps perceptually similar images to nearby points in a reduced dimensional space) and then use the $\ell_2$ distance in this encoded space. 
If the mean distance falls below a chosen threshold, we flag this query as a potential attempt at generating an adversarial example.
We choose the detection threshold so that benign use of the classifier is not flagged. In particular, we set the threshold so that if the entire training set were to be randomly streamed as queries, $0.1\%$ of the training set would be flagged
as attacks (i.e. the false positive rate would be $0.1\%$). 
\footnote{In practice, a lower false positive rate may be necessary. 
However, some existing defense research for detecting adversarial examples sets the false positive rate at approximately $5\%$ \cite{feature-squeezing}. Our value is thus $50\times$ lower than prior work.}
After an attack is detected, the buffer containing the previously saved queries for that user can then be cleared. Moreover, in response to the attempted attack, the user may then be banned from the service either immediately, or after a random number of subsequent queries, in order to reduce the attacker's knowledge of when exactly their attack was detected. A diagram of our scheme is shown in Figure \ref{fig:scheme}.

\subsection{Similarity Encoder} \label{sim-encoder}
A key question in the design of our method is the metric to use for the
$k$-nearest-neighbor search.
Naively, we might imagine choosing a simple metric---for example, the $\ell_2$ distance between two images.
However, using such a simple method has two drawbacks:
\begin{enumerate}
    \item Simple metrics, such as $\ell_2$, do not accurately capture distance in adversarial situations and are too easy for an attacker to evade.
    A small rotation or translation in pixel space can cause dramatic changes according to the $\ell_2$ norm, which experimentally we find allows an adversary to evade detection.
    
    \item Also, the $\ell_2$ distance requires storing an entire copy of every queried image. This could pose a significant cost to the hosting service in storage costs, and storing user queries for longer than strictly necessary introduces potential privacy risks.
\end{enumerate}

To increase the security of our scheme, we use perceptual similarity for nearest-neighbor search, as adversarial attacks involve generating new images that are perceptually similar to the original image.
To measure the perceptual similarity of two images, we train a deep neural network to encode images into a lower-dimensional space of dimension $d$, such that similar images are mapped to similar points in the encoded space.
For example, for a given picture of a dog, after rotating or translating the image slightly, the perceptual content of the image is still the same (i.e., the same dog), and we train the encoder so that both of these images have similar $d$-dimensional representations.

This construction resolves both of the difficulties identified earlier. 
By design, small modifications to an image are less likely to cause dramatic increases in encoded-space $\ell_2$ distance.
Further, because the encoded space is much smaller than the total image size, this allows us to save on storage costs.

\textbf{Encoder Setup \& Training.}
We represent the encoder $E(\cdot)$ as a neural network mapping images $x \in \mathbb{R}^{h\cdot w \cdot c}$ to an encoded space $e \in \mathbb{R}^d$ of dimension $d$.
As described, the objective of this encoder is to map visually similar inputs $x,\tilde{x}$ to encodings $e=E(x)$, $\tilde{e}=E(\tilde{x})$ that are similar under $\ell_2$ distance, so that $\lVert e-\tilde{e}\rVert_2$ is small.

To achieve this we train the similarity encoder neural network with a contrastive loss function \cite{bell15productnet}. 
Specifically, we consider two pairs of images. Pair 1 consists of $x_i$, an image drawn from the training set, and $x_p$, a ``positive'' image perceptually similar to $x_i$. Pair 2 consists of a different training image $x_j$, along with
a negative example $x_n$, an image \emph{not} perceptually similar to $x_j$.
The contrastive loss for their encodings  $(e_i, e_p), (e_j, e_n)$ is
\[L(x_i, x_p, x_j, x_n) = {\|e_i - e_p \|}_2^2 + \max(0, m^2 - {\|e_j - e_n\|}_2^2).\]
The first term encourages similar encodings for positives and the second term encourages different encodings for negatives by
penalizing encodings less than a certain margin $m$ apart for negatives.

\subsection{Experimental Setup}
We evaluate our defense on the CIFAR-10 dataset: a collection of 60,000 low-resolution ($32\times32$) color images drawn from 10 classes,
so that each image is in $[0,1]^{32\times32\times3}$.
We choose this dataset for three reasons. First, CIFAR-10 is the most popular dataset for studying adversarial examples \cite{obfuscated-gradients}. Second, defenses on ImageNet have thus far proven to be far beyond our current capabilities \cite{ALPBreak}, and no known neural network defense remains robust to attack. Finally, training a network for CIFAR-10 is significantly less computationally expensive than for ImageNet, allowing for us to perform a wide range of experiments.

We train a ResNet classifier \cite{ResNet} on the CIFAR-10 dataset for 100 epochs with Adam \cite{ADAM} on a 1080 Ti GPU with Keras \cite{Keras} and TensorFlow \cite{Tensorflow}, achieving $92\%$ test accuracy. 

\begin{figure}
  \begin{center}
  \centering
  \includegraphics[width=0.45\textwidth]{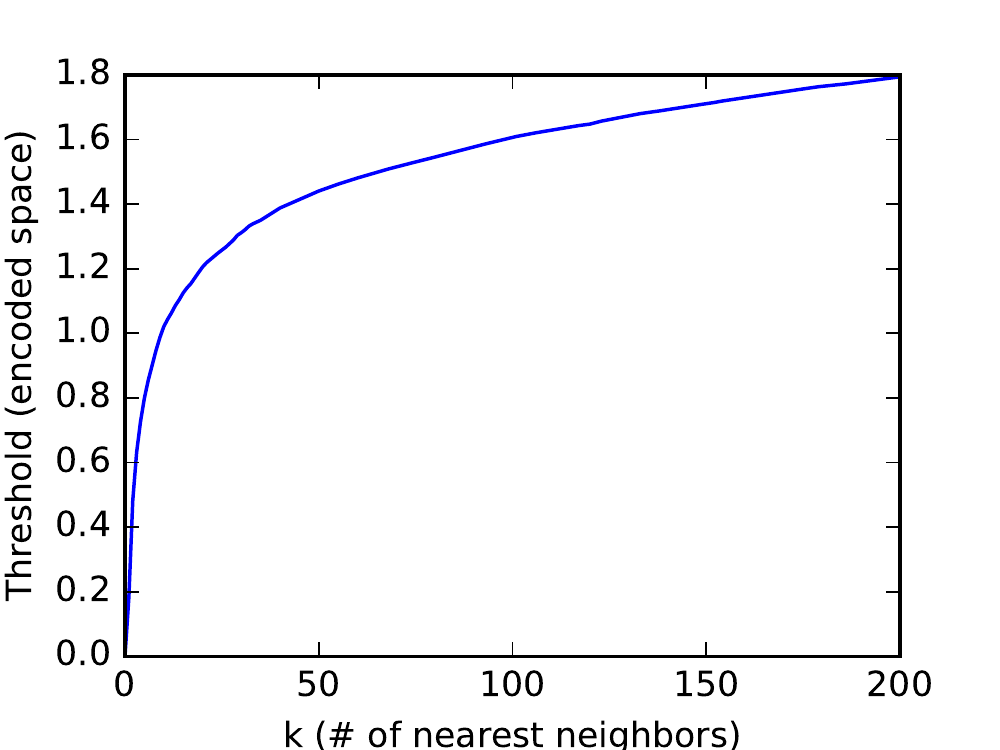}
  \caption{The mean $k$-neighbor distance (in encoded space) of the $0.1\%$ percentile of the CIFAR-10 training set as a function of $k$.
  We select the threshold $k$ so that it is large enough to support a wide margin, but not so large as to be computationally prohibitive.}
  \label{fig:kdist}
  \end{center}
\end{figure}

\subsection{Encoder Training and Threshold Selection}
To train the similarity encoder for the scheme, we follow prior work \cite{bell15productnet} and  initialize our encoder with the same architecture and weights as a network trained to classify the desired images.
For CIFAR-10, we first train a three-layer CNN (architecture given in the appendix) for 100 epochs using data augmentation and reach a validation accuracy of $76\%$. Then we remove the final layer and replace it with a new encoding layer of dimension $d = 256$. We found a margin of $m = \sqrt{10}$ experimentally resulted in the best encodings. A more complex architecture might yield stronger results, but
to keep our initial design simple we use this CNN.

We fine-tune this neural network (with the replaced last layer) to minimize the contrastive loss function described in Section \ref{sim-encoder}, with a learning rate $\alpha = 1e-4$, momentum $\mu = 0.9$, and a batch size of $b = 32$. To generate a batch of positive pairs for training, we sample $b$ images from the training set; then, a random image transformation (that should retain the perceptual content of the image) was selected and applied to each of the $b$ images (similar to traditional data augmentation used for network training). For negative pairs, we sample $b$ pairs ($2b$ images total) of different images. The transformations used are enumerated in Section \ref{section:transformations}.

\textbf{Parameter Selection.}
The choice of $k$, the number of neighbors, affects the effectiveness of our scheme.  Large values can potentially improve the effectiveness of our scheme at detecting attacks (since larger $k$ allows a larger threshold while maintaining 0.1\% false positive rate, thus forcing the attacker's images to be very different to avoid detection).
However, $k$ is the minimum number of queries before our defense could possibly flag a possible attack, so smaller values of $k$ can potentially enable faster detection of attacks; and smaller values of $k$ reduce the computational cost of calculating the $k$-neighbor distance.
Therefore, a smaller $k$ should be preferred whenever possible.

We show in Figure \ref{fig:kdist} the threshold needed to ensure a $0.1\%$ false positive rate, as a function of $k$.
Larger values of the threshold constrain the attacker more sharply if the attacker wishes to avoid detection, so the larger the threshold the better.
We can see that this threshold increases sharply until about $k = 50$, where the distance begins to plateau and marginally continues to increase as $k$ increases. Therefore, we select $k = 50$.

\section{Non-Adaptive Evaluation}
Having described our defense proposal, we begin by demonstrating that it has at least some potential utility: it effectively detects existing (unmodified) black-box query attacks.
While there are many black-box (hard-label) attacks, they fall roughly into two categories:
\begin{itemize}
    \item \textbf{Gradient estimation} attacks at their core operate like standard
    white-box gradient-based attacks (as described in Section \ref{subsection:preliminaries}).
    However, because they do not have access to the gradient, these types
    of attacks instead estimate the gradient by repeatedly querying the model.
    \item \textbf{Boundary following} attacks, in contrast, first identify the
    decision boundary of the neural network, at a potentially far-away point,
    and then take steps following the boundary to locate the nearest point on
    the boundary to the target image.
\end{itemize}
We evaluate against one representative attack from each category.

\subsection{Attack Setup}
For each attack studied, we use the \emph{targeted} variant, where the adversary generates an adversarial example chosen so that the resulting adversarial example $x$ is classified as a target class $t$ and is within a distance $\epsilon$ of an original image $x$.
The original image and target class are chosen randomly.
We call an attack successful if the $\ell_{\infty}$ distortion is below $\varepsilon=0.05$.
While most white-box work on CIFAR-10 considers the smaller distortion bound of $\varepsilon=0.031 \approx 8/255$, we choose this slightly larger distortion because black-box attacks are known to be more difficult to generate and so we give the adversary slightly more power to compensate.

\textbf{NES}
\cite{NES} is one of the two most prominent gradient-estimation attacks (along with SPSA \cite{SPSA}).
It estimates the gradient at a point by averaging the confidence scores of randomly sampled nearby points, and then uses projected gradient descent \cite{madry2017towards} to perturb an image of the target class until it is sufficiently close to the original image. In the hard label case, the confidence score for a point is approximated by taking a Monte Carlo sample of nearby points, and then computing the score for a class as the fraction of nearby points with that class.

\textbf{The Boundary Attack}
\cite{BoundaryAttack} was the first attack to propose following the decision boundary to generate black-box adversarial examples.
Since its publication, there have been multiple proposals to improve this attack \cite{RED, Boundary++}. We evaluate our defense on the vanilla boundary attack; the other attacks are more query efficient, but at their core still perform the same operation. To compare directly with the NES attack, we use $\ell_{\infty}$ distance with the boundary attack (instead of the usual $\ell_2$ distance). 

\subsection{Results}
We run the default, unmodified implementations of each attack against our scheme and find that they can be detected. The results are presented in Table \ref{table:non-adaptive-results}. (Attack-specific parameters for NES are given in Section~\ref{NES}.)  An attack is considered successful if an adversarial example is found within an $\ell_\infty$ distortion of $\epsilon = 0.05$ from the original image, with the correct target class. We terminate each attack as soon as it finds such an example. Each attack does eventually succeed at a high rate, but is detected frequently with at least 200 detections on average. Thus, an attacker would need to create at least 200 accounts in order to generate a single adversarial example with these attacks. This demonstrates that our query sequence based scheme can reliably detect existing black box attacks.
\begin{table}
	\begin{tabular}{llrrrl}
	\toprule
                  &   Attack              & Success Rate & Num. Queries  & Detections \\
                  \midrule
\multirow{3}{*}{} & NES             & 100\%               & 325,200$\pm$153,300         &     6,377                          \\                                       
                  & Boundary        & 100\%               & 14,720$\pm$8,923            &     288   \\
                  \bottomrule
		\end{tabular}
  \caption{\textbf{Success rate of unmodified attacks on a neural network protected with our scheme.} While attacks succeed with $100\%$ probability, the attacks trigger hundreds to thousands of detections.}
  \label{table:non-adaptive-results}
\end{table}

\section{Query Blinding}
While showing that our proposed defense can detect existing attacks is a useful first step, it is not sufficient for a complete evaluation. We must also evaluate whether our defense can detect future attacks.
Doing this requires developing \emph{adaptive attacks} specifically designed to bypass the defense proposal \cite{carlini2019evaluating}.

Thus, we introduce \emph{query blinding}%
\footnote{We call this \emph{query blinding} because of its similarities to \emph{blind signatures} \cite{chaum1983blind}.}: a general strategy which can be used to hide the query sequence from the defender.
Query blinding is the most effective attack strategy we have found to date against our scheme.
At its core, the objective of a query blinding attack is to learn the value of $f(x)$, for some specific $x$, without actually revealing the example $x$ to the defender.
We define two functions: a randomized \emph{blinding function} $b(x; r) = \{x'_0, x'_1, \dots, x'_n\}$ that maps from one example to a set of modified examples so that $\lVert{}x'_i-x\rVert{} \ge \varepsilon$, and
a \emph{revealing function} $r(f(x'_0), f(x'_1), \dots, f(x'_n))$ that is designed to estimate $f(x)$ from the classifier outputs.
For the majority of this paper we restrict ourselves to the case where $\lVert b(x) \rVert=1$.
In the appendix we give an example of a more sophisticated blinding function that deviates from this simplifying assumption.

\subsection{Image Transformations} \label{section:transformations}
Image processing transformations, such as image translation and brightness adjustment, are natural and readily available blinding functions. Let $x$ be the image that an attacker would like to query the model for, $f(x)$ be the model's output for the query, and $T_c(x; r)$ be a randomized image processing transform (e.g., by rotating it or shifting it by a random amount $c$);
then we set $b(x;r) = \{T_c(x;r)\}$. 
Because the purpose of our blinding function is to fool the query-detector by transforming a sequence of queries which are pairwise similar to a sequence of queries which are not, we would like the distortion between the original image and the transformed image to be large.
For example, for a 3 x 32 x 32 CIFAR-10 image, adjusting the brightness (i.e., individual pixel values) by adding just $0.05$ to each pixel increases the $\ell_2$ distortion by $0.05 \times \sqrt{3 \times 32 \times 32} = 2.77$. 
Critically, after distortion, these transformations still retain the primary content of the image, and a model with high accuracy should produce relatively similar outputs for the original and transformed images, so the corresponding revealing function for an image processing transformation is simply $r(f(x')) = f(x')$.
We examined eight possible transformations:
\begin{itemize}
\item \textit{Uniform Noise}: add uniform noise to the image, where the noise is drawn from a uniform distribution $c \sim U(-r, r)$. 
\item \textit{Translate}: translate the image horizontally and vertically by $c_h$ and $c_v$ fractional pixels, where $c_h$ and $c_v$ are sampled randomly from a uniform distribution $c_h \sim U(-r, r)$ and $c_v \sim U(-r, r)$, using bilinear interpolation and filling in empty space with zeroes. 
 
\item \textit{Rotate}: rotate the image $c\pi$ radians, where $c$ is sampled randomly from a uniform distribution $c \sim U(-r, r)$, using bilinear interpolation and filling in empty space with zeroes. 

\item \textit{Pixel-wise Scale}: scale all pixels by the same factor $c$, sampled randomly from a uniform distribution $c \sim U(1 - r, 1 + r)$. 

\item \textit{Crop and Resize}: crop the image to box coordinates of $[c, c, 1 - c, 1 - c]$ and then resize the image to its original size (box coordinates $[0, 0, 1, 1]$) using bilinear interpolation, where $c$ is sampled randomly from a uniform distribution $c \sim U(0, r)$. 

\item \textit{Brightness}: adjust the brightness of the image by adding to each pixel the same constant $c$, where $c$ is sampled randomly from a uniform distribution $c \sim U(-r, r)$. 

\item \textit{Contrast}: adjust the contrast of the image by a random contrast factor $c$, where $c$ is sampled randomly from a uniform distribution $c \sim U(r, 1)$. 

\item \textit{Gaussian Noise}: add Gaussian noise to the image, where the noise is drawn from a Gaussian distribution $c \sim r \times \mathcal{N}(0, I)$. 

\end{itemize}
These transformations are also used when training the similarity encoder (Section \ref{sim-encoder}).

\subsection{Auto-Encoder Attack}
We also consider attacks that involve \emph{learning} the blinding function.
Specifically, we train an auto-encoder neural network $\alpha(x)$. 
Normally, auto-encoders are trained so that $\alpha(x) \approx x$. 
In our case we instead train a randomized auto-encoder $\alpha(x; r)$ to satisfy two properties:
(1) $\lVert \alpha(x; r_1) - \alpha(x; r_2) \rVert|_2$ is large, but
(2) $f(\alpha(x; r)) \approx f(x)$.
Satisfying property (1) ensures that the augmented image will evade detection by the encoder, 
while property (2) ensures that the actual classification of the image will remain unchanged.

Specifically, we train the auto-encoder to minimize the loss
\[\ell(x) =   H(f(\alpha(x, r)), f(x)) - c \cdot \min(\lVert\alpha(x; r_1)-\alpha(x; r_2)\rVert_2^2, d^2) \]
where $H(\cdot)$ is the cross-entropy loss between the two distributions, $c$ is a constant that controls the relative importance of the two loss terms, and $d$ is a constant that sets the desired $\ell_2$ distance between transformed examples.

We train the auto-encoder with stochastic gradient descent for $10$ epochs on the CIFAR-10 training data.
In order to ensure that we are not ``cheating'' by training on the exact function $f(\cdot)$ which we will be attacking, we train a new classification neural network $f'(\cdot)$ on $10\%$ of the CIFAR-10 training data.
In practice, we set $c=1$.

To determine the threshold $d$, we try values between $2$ and $20$ and pick the one that is most effective at fooling the detector. We found that in practice $d=10$ is well-balanced between being big enough so the detector is fooled, but not so big that $f(\alpha(x;r))$ is substantially different from $f(x)$.

\subsection{Increasing Attack Diversity}
While the above query-blinding attacks generate blinded images at a large distance from the original, if we are unlucky they might generate two blinded images which are nearly identical.
This is true especially when we generate a large number of transformed images $x_i = b(x, r_i)$. By the birthday paradox, we should expect that as we generate increasingly many images, an increasingly large fraction will be similar to each other, which might cause the attack to be detected.
We improve our attacks so they avoid querying multiple blinded images that are too similar.

For simplicity of analysis, assume for the moment that our defense relied exclusively on the $\ell_2$ distance between images being less than some threshold $\tau$ to detect attacks.
After the attacker has made queries ${q_0, q_1, \dots, q_{j-1}}$, he could check whether or not making the query $q_j = b(x; r_j)$ would be detected as an attack against his own history.
If it would be, he could simply re-sample a new value $r_{j'}$ and generate a new candidate query $q_j = b(x; r_{j'})$ until he obtained a sample that is above the detection threshold.
This greedy approach is simple and reasonably effective.

One can do even better by filtering all queries in advance.
Begin by generating a large number of candidates $x_i = b(x, r_i)$.
Construct a graph $G$ where each candidate $x_i$ is a node, and connect node $i$ to node $j$ if the distance between $x_i$ and $x_j$ is less than the threshold $\tau$ (i.e., querying both would result in a detection).
Formally, $G=(V,E)$ where $V = {1, \dots N}$ and $E = \{(i,j) : \lVert x_i - x_j \rVert < \tau\}$.
If the defender uses $k=2$, identifying the maximum subset of examples that would not cause a detection by a $\ell_2$-based detector reduces to finding the maximum independent set of this graph $G$.
While the maximum independent set in general is NP-hard to approximate, we find that simple approximation algorithms identify sub-graphs with a cardinality on average twice as large as greedy querying.

We use these methods to improve our query blinding attacks.
Even though our defense actually uses an encoder and does not rely directly on $\ell_2$ distance, we find that this approach still reduces the number of detections by promoting attack sample diversity.

\section{Adaptive Attack Evaluation} \label{section:attacks-description}
Given that our proposed defense effectively prevents existing black-box attacks, we now study whether or not it can prevent more sophisticated attacks.
We find that while it is possible to degrade the effectiveness of the defense, we can not defeat it completely.
We study three categories of attacks: gradient attacks (specifically, variants of NES with various kinds of query blinding), boundary-following attacks (specifically, variants of the boundary attack with query blinding), and hybrid attacks (which combine gradient attacks with a surrogate model).

\subsection{The NES Attack} \label{NES}
We use the NES attack as one starting point for attacking our scheme. To generate a targeted adversarial example a given input $x$, NES generates an adversarial example by seeding it with an image $x'$ of the target class (i.e., that is already adversarial). NES then uses projected gradient descent to slowly reduce the distortion between this image (which is already the target class) and the original example $x$ until it is within $\epsilon$ of the original image, while still being classified as the target class.

In order to estimate the gradient at any given location $x$, NES uses finite differences on a random Gaussian basis. Full details can be found in \cite{NES}, but simplified, the gradient is estimated by: (1) sampling $n$ instances of Gaussian noise $\delta_1,\dots,\delta_n \sim \mathcal{N}(0, 1)$  and adding them each to $x$ as $\theta_i = x + \sigma \delta_i$ to generate $n$ basis points, (2) for each basis point $\theta_i$, estimating the confidence scores at $\theta_i$, (3) estimating the gradient at $x$ using these estimated confidence scores and then taking a step in the direction of the estimated gradient.
The confidence score at $\theta_i$ is estimated by querying the labels for $s$ points near $\theta_i$ chosen randomly from a sampling ball of $l_{\infty}$ radius $\mu$ and computing the proportion of each class as the estimate for that class's confidence score.
The default attack parameters for NES are $\sigma = 0.001$, $n = 4$, $s = 50$, $\mu=0.001$, and learning rate = $0.01$ \cite{NES}; we consider below how to adjust them.

An attacker aware of our defense may attempt to modify the NES attack in order to increase the distance between queries. 
We explore two natural modifications to the NES attack that an attacker could make: adjusting the NES parameters, and query blinding.

\subsubsection{Parameter Tweaking}
\begin{figure}
  \begin{center}
  \includegraphics[width=0.45\textwidth]{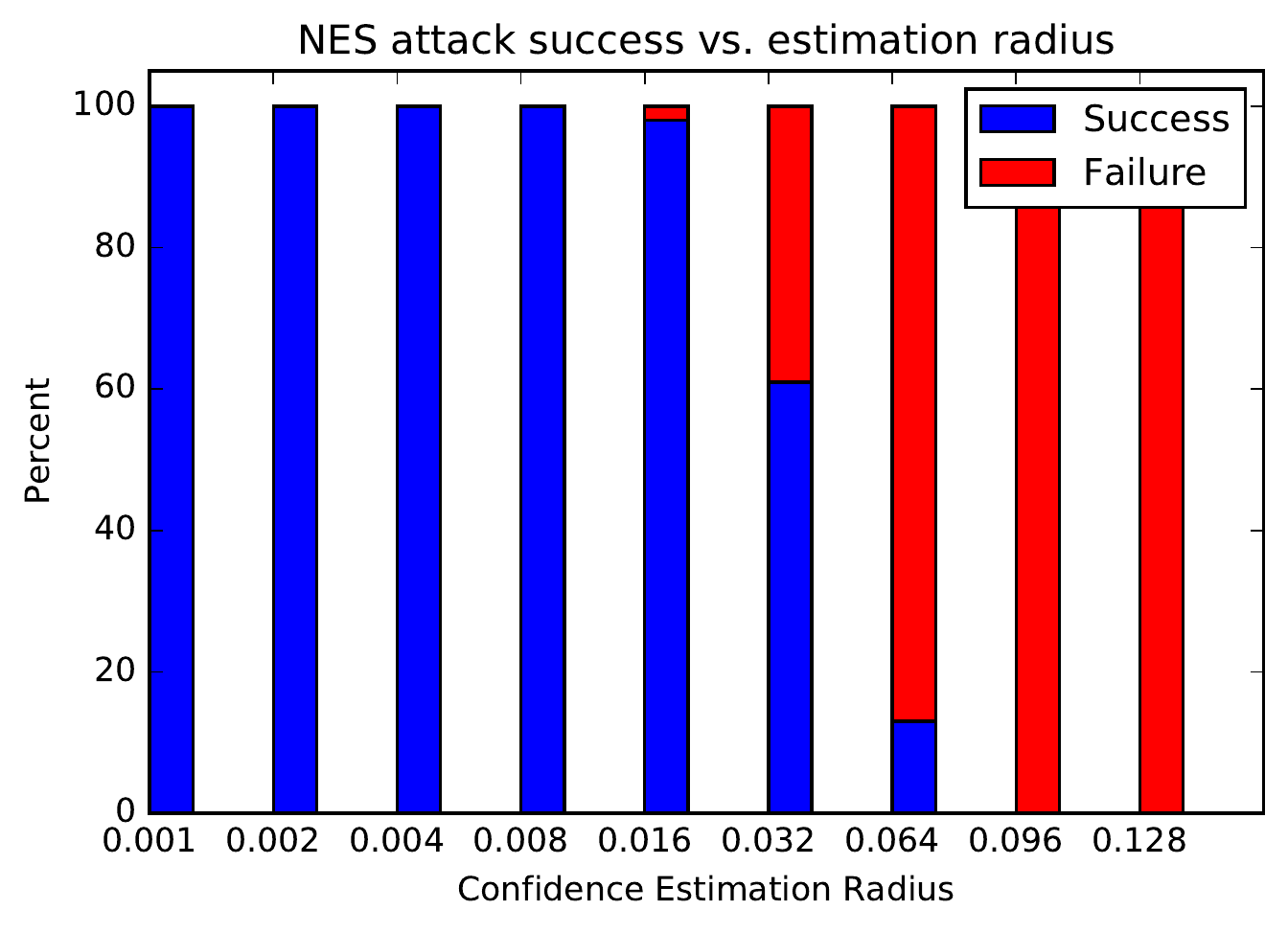}
  \caption{NES attack success rate, as we vary the sampling radius $\mu$. For each value of $\mu$, we ran the NES attack was run 100 times; the attack was considered successful if it found an adversarial example (of a randomly chosen target class and within $\epsilon = 0.05$) within 5 million queries.}
  \label{fig:nes-radius}
  \end{center}
\end{figure}

One natural modification is to adjust NES's parameters to make it harder for our defense to detect the attack.
The attacker could increase $\mu$, the radius of the sampling ball used when sampling points to estimate the confidence score for an image. The original version of the NES attack uses a radius of $\mu = 0.001$, but in experiments we found that the attack is also reasonably successful with radii up to $\mu = 0.064$ (see Figure \ref{fig:nes-radius}).
When we set $\mu = 0.064$, the 50-nearest-neighbor distance between an image and the $s$ sampled points is on average $2.32$, significantly larger than the 50-nearest-neighbor distance of just $0.032$ if we set $\mu = 0.001$. 

The attacker could also increase $\sigma$ so that the sampled Gaussian basis points $\theta_i$ are further apart. Additionally, the attacker could decrease $s$ so that fewer queries near $\theta_i$ are generating when estimating confidence scores.  We found that the attack remains reasonably successful even when $s$ is reduced to $s=2$, and the attack becomes  harder to detect, so we used $s=2$ in our experiments.  Figure~\ref{table:NEStransformResultsExtended} in the appendix reports on the effect of modifying this parameter.

\subsubsection{Query Blinding}
The second natural modification is for an attacker to use query blinding to transform each query, as described previously.
In particular, we modify the confidence score estimation procedure (step 2 from the previous attack description) to sample $s$ points using the different transformations listed in Section \ref{section:transformations} instead of sampling from a $\ell_\infty$ ball of uniform radius. The parameters for each transformation are normalized so that the expected $\ell_2$ distortion from each transformation is equal to $2.32$; exact values are given in Appendix~\ref{sec:transform-params}.
We selected a constant of $2.32$ to match the $\ell_2$ distortion of setting $\mu = 0.064$. (Note that $\mu$ is now only applicable when using the original strategy of sampling from a ball of $\ell_\infty$ radius $\mu$).
We used the same parameters when training the similarity encoder.


\begin{table*}
	\begin{tabular}{llrrrrl}
	\toprule
                  & Attack              & Attack Success Rate & Num. Queries  & Detections   & Detections with $\ell_2$ detector\\
                  \midrule
\multirow{8}{*}{\rotatebox{90}{Low Distortion}} & NES (query blinding: uniform noise)   & 1\%                 &  15,700$\pm$0          & 308           & 308                               \\
                  & NES (query blinding: translate)       & 4\%                 &   6,710$\pm$400      & 131           & 131                               \\
                  & NES (query blinding: rotate)          & 7\%                  &  10,200$\pm$684      & 198           & 199                               \\
                  & NES (query blinding: scale)           & 27\%                 & 12,600$\pm$882       & 246           & 246                                \\
                  & NES (query blinding: crop and resize) &  7\%               & 6,130$\pm$247        & 119           & 119                                \\
                  & NES (query blinding: brightness)      & 55\%               & 13,500$\pm$780       & 264          & 263                                \\
                  & NES (query blinding: contrast)        & 23\%               & 11,200$\pm$777       & 219           & 219                                \\
                  & NES (query blinding: Gaussian noise)  & 0\%                  & n/a           & n/a           & n/a \\
                  \midrule
\multirow{3}{*}{\rotatebox{90}{High}} & NES (query blinding: brightness)   & 43\%             & 24,500$\pm$2,630          & 481               & 60   \\
                  & NES (query blinding: scale)       & 42\%              & 25,700$\pm$3,000          & 504               & 88    \\
                  & NES (query blinding: contrast)       & 37\%           & 19,100$\pm$4,240          & 375               & 85   \\
                  \midrule
                  & NES (query blinding: auto-encoder)      &       76\%         &   13,400$\pm$8,400           &    97 & 12                   \\
                  \bottomrule
		\end{tabular}
\caption{The success rate of NES attacks with query blinding. The fourth column shows the average number of times the attack is detected my our defense before a single adversarial example is generated (i.e., the number of accounts an attacker would need to defeat our defense). The fifth column shows the average number of times the attack is detected, if we use a simplified defense with $\ell_2$ distance on images instead of the similarity encoder. Simple transformations, such as rotation or contrast adjustment, are still detected by our defense, even when the distortion introduced is large. The auto-encoder attack is more effective but is still detected by our defense.  These results show the value of the similarity encoder; it makes our defense significantly more effective against several attacks.}
  \label{table:NEStransformResults}
\end{table*}

When running the NES attack, each time we query the classifier we preprocess the image with one strategy. We use default parameter values for the NES attack, except we set $s=2$ to make it harder to detect.
Table \ref{table:NEStransformResults} shows the effectiveness of different transformations. For some transformations, like uniform and Gaussian noise, the NES attack fails completely. However, the NES attack works even better with brightness and pixel-scale transformations than the original confidence estimation procedure of uniform noise. This suggests that estimating the confidence score for an image may be more accurate with certain image transformations than others. 

For all transformations, each attack will trigger at least one hundred detections (on average), so our defense is effective at detecting these query blinding attacks. The exact attacker cost corresponding to this number of detections is quantified further in Section \ref{section:attack-economics}. 


For this level of transformation distortion, the similarity encoder offers little benefit over $\ell_2$ distance on images. This is understandable, as for $k = 50$, the $\ell_2$ detection threshold is $\delta = 5.069$ when using $\ell_2$ distance on images, which is greater than the $\ell_2$ distortion of $2.32$ introduced by these transformations. We also evaluated against attacks that use transformations that introduce a greater distortion. Specifically, we increase $\sigma$ to $\sigma = 0.01$ (the highest value found experimentally such that the attack still succeeds reasonably often) and increase the distortion parameter for each transformation so that the expected $\ell_2$ distortion introduced by the transformation is equal to $5.10$ (see Appendix~\ref{sec:transform-params}). Results for these higher-distortion attacks are  shown in Table \ref{table:NEStransformResults}.

%

\begin{figure}
    \centering
    \vspace{-3.5em}
    \includegraphics[clip, trim=0in 9in 1.5in 0in]{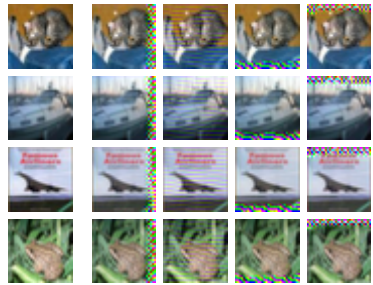}
    \caption{For each image in the test set (left) we show four possible images transformed by our auto-encoder (right) that have the same classification label but have high $\ell_2$ distance from each other.}
    \label{fig:transformed_examples}
    \vspace{-1em}
\end{figure}

Against these higher-distortion transformations, the similarity encoder is more effective at detecting attacks compared to $\ell_2$ distance on images, resulting in 6--8$\times$ more detections. As explained in Section \ref{section:attack-economics}, this significantly increases the cost to the attacker.  This demonstrates that the similarity encoder significantly increases robustness against adaptive query blinding attacks.
The number of detections increases when using higher-distortion transformations vs lower-distortion transformations; this is because the attack needs more queries on average to succeed (due to less precise gradient estimates), resulting in more opportunities to detect the attack.

The learned auto-encoder is more effective than simple image transformations. (We give examples of the auto-encoded images which are sent to the classifier in Figure~\ref{fig:transformed_examples}.)
The auto-encoder is particularly effective at defeating a defense that uses $\ell_2$ distance on images, as it is trained to map an image to a blinded version with large $\ell_2$ distortion.
However, our full defense with a similarity encoder remains able to detect the auto-encoder attack.
Future work may be able to improve the auto-encoder attack by training the auto-encoder to
fool surrogate similarity encoders.



\subsection{The Boundary Attack}
The boundary attack \cite{BoundaryAttack} is a gradient-free attack that starts with an image of the target class and then makes steps alternating between moving the image along the decision boundary (while remaining adversarial) and steps which move towards the original image.
In more detail, the boundary attack alternates between the two following operations:
\begin{itemize}
\item \textbf{Inwards step.} Given the current proposed adversarial example $x'_i$, we take a small step $\varepsilon$ in the direction of the original image $x$ and let $x'_{i+1} = x'_i + \varepsilon \cdot (x - x'_i)$.
 \item \textbf{Orthogonal (boundary-following) step.}
 Starting from the current adversarial example $x'_i$, take a step along a random direction $r$ orthogonal to $x-x'_i$, subject to the constraint that $f(x'_i+r)=f(x'_i)$.
\end{itemize}

\subsubsection{Parameter Tweaking.} 
An attacker may also use different variants of the boundary attack in an attempt to avoid detection by our scheme.
First, the attacker could increase the step sizes to be large enough so that they are not detected by our similarity encoder; in our experiments, we found that the boundary attack no longer converges after this change.
Second, instead of sampling random directions from a Gaussian distribution, as in the original attack, an attacker could sample from Perlin noise. This substitution was shown to increase success rate with a limited number of queries (fewer than 15,000 queries) \cite{BiasedBoundary}.
However, we find that this strategy is not effective at the higher number of queries allowed in our evaluation (beyond 20,000 queries). This is likely due to the decreasing marginal improvement of this Perlin substitution as queries increase, an effect observed in \cite{BiasedBoundary}.

\subsubsection{Query Blinding.}
We also evaluate the effectiveness of a boundary attack with query blinding. In particular, we preprocess all queries made by the boundary attack with the transformation that worked best with the NES attack (brightness).

\subsubsection{Results.}
Table \ref{table:results} shows our ability to detect these versions of the boundary attack. We allow each attack trial to make up to 200,000 queries, the same number of queries made by the best NES attack variants. Perlin noise does not perform better than the original boundary attack, likely due to the observed decreasing utility of the Perlin noise at these higher query numbers \cite{BiasedBoundary}. Preprocessing queries with the brightness transform is also ineffective: it does not decrease the number of detections while significantly decreasing the boundary attack's success rate. We found that the success rate decreases because the boundary attack adjusts its step size according to the local geometry of the boundary (estimated from past queries); the preprocessing causes the attack to misapproximate the local boundary geometry and reduce the step size too drastically, causing the attack to converge prematurely before finding a point within distance $\epsilon$ from the original image $x$ and fail to make further progress. 
Consequently, the original boundary attack is more effective than any of the variants we studied.


Our defense is effective at detecting the boundary attack: an attacker would need to create 200 accounts to create a single adversarial example.

\begin{table*}
\begin{tabular}{llrrrl}
\toprule
                  &   Attack              & Attack Success Rate        & Num. Queries & Detections \\
                  \midrule
                  & Boundary attack (normal)            & 100\%    & 14,700$\pm$892            &     288         \\
                  & Boundary attack (Perlin noise)              &  40\%  & 31,100$\pm$976            & 609                       \\
                  & Boundary attack (query blinding: brightness)          &  2\%   & 24,100$\pm$1,470             & 473                       \\
                  & Boundary attack (query blinding: auto-encoder)          &        61\%        &  34,200 $\pm$ 8,000        &    240         \\
                  
                  \midrule
                  & Hybrid surrogate attack, untargeted  &  14\%        & 8,192$\pm$0                   & 82 \\
                  \bottomrule
\end{tabular}
\vspace{.5em}
\caption{Effectiveness of the boundary attack and the hybrid surrogate attack. Our defense is able to detect all of these attacks.}
  \label{table:results}
\end{table*}

\subsection{Hybrid Query Based Surrogate Attack}
Gradient-estimation query based attacks by design necessarily make a sequence of queries which are all highly self-similar, and our detection defense is especially well suited to these types of attacks. Thus, we now consider an alternate attack strategy.
At a high level, we will construct a locally-similar surrogate model \cite{papernot2016transfer} and then perform a gradient-based attack on this model and
transfer the attack.
Specifically, for an image $x$, we randomly sample $n$ points, $x_\text{far} = x + \delta, \ \delta \sim U(D, 1)$, where $D$ is large enough that querying these points is unlikely to trigger the defense, and we query the sampled points for their labels. To ensure that some training points of $x$'s class are present, we also sample $m$ points, $x_\text{near} = x + \delta,\  \delta \sim U(0, d)$,  where $d$ is small, but we do \emph{not} query these points; instead, we assume their labels are of the same class as $x$, which is normally the case unless $x$ is very close to a decision boundary. We train a surrogate model on these $n + m$ points, and then run the white-box attack from Carlini and Wagner \cite{Carlini} with high confidence ($\kappa = 100$), in order to generate an adversarial example that hopefully transfers. In our scheme, the similarity encoder is not made public, so the attacker would not know the exact value of $D$ to use, but in our experiments we artificially give the attacker the extra power to know the optimal $D$, to give the attacker every possible advantage.


In our experiments, we found that targeted version of this attack did not succeed, as it was difficult to find enough points with the target class through random sampling.
Therefore, for the evaluation of this attack, we focus on untargeted attacks and exclusively target the next-most-likely class, instead of a random class.
We selected $D$ as $D = 0.09$, by experimentally finding the necessary value such that samples would on average have a higher similarity encoding distance than the threshold, and $d = 0.01$. We use a three layer CNN architecture (described in the appendix) for the surrogate model, and trained it for 50 epochs per attack.
Table \ref{table:results} shows the effectiveness of this attack. This attack is able to succeed at a modest but non-trivial rate, while having 3--8$\times$ fewer detections compared to the other attacks. This indicates that there may be potential for constructing query-based attacks that use information from surrogate models to make the attack stealthier; but we have not found a concrete attack that can completely defeat our defense. We leave it to future work to explore improving such attacks (e.g., informed instead of random sampling).

\subsection{Attacking the Similarity Encoder}
An attacker might try to defeat our scheme by fooling the similarity encoder,
e.g., generating a blinded version $x'$ of a query $x$ such that the similarity encoder
considers $x,x'$ to be dissimilar, yet they have the same classification.
Because the similarity encoder is kept secret in our defense, there is no way to perform a
white-box attack to construct such blinded queries.
There is no direct way to mount a black-box query attack, either: there is no way to supply
an image $x$ and learn its encoding, or submit a pair of images $x_0,x_1$ and learn whether they have similar encodings.

This leaves only side-channel attacks on the similarity encoder.
For instance, an attacker could create a new account, submit a batch of $k+1$ images, and observe whether the account is cancelled or not; this reveals whether the $k+1$st image was similar to the previous $k$.
However, such an attack would be very slow.
Each $k+1$ queries reveal only a single bit of information about the similarity encoder.
As constructing a surrogate model or mounting query-based attacks typically require tens or hundreds of thousands of examples, such an attack would require creating an infeasible number of fake accounts.
Information-theoretically, to maximize the amount of information revealed per query, the optimal strategy is to issue $k=50$ queries, then issue subsequent queries chosen so that each has about a $p\approx 1/18$ chance of triggering detection and observe which one causes the account to be cancelled; this reveals about 5.6 bits of information per account and issues about 68 queries per account (on average).
We expect that it would still require thousands of fake accounts, so is unlikely to be effective.

If necessary, a defender could further reduce the amount of information revealed by the side channel by delaying account cancellation.
For example, if the defender waits until the number of queries reaches the next power of two before banning an attacker's account, and if no user makes more than $2^{20}$ queries, then each account cancellation reveals at most $\lg 21 \approx 4.4$ bits of information about the encoder, at the cost that direct attacks might require only half as many fake accounts.
The parameters could be optimized: for instance, if accounts are cancelled only at queries numbered $\lceil 50 \times 1.1^i \rceil$ for $i=1,2,\dots$ and if no user makes more than $2^{20}$ queries, then each account cancellation reveals at most $\lg \log_{1.1} (2^{20}/50) \approx 6.7$ bits of information about the encoder, and the delayed cancellation allows the attacker to make at most 10\% more free queries; thus our best attack (NES with auto-encoder query blinding) would still require $97/1.1 \approx 88$ fake accounts with this side-channel defense in place.

\section{Economics of Performing an Attack} \label{section:attack-economics}
Assuming our defense was in place, what would the economics look like
for an attacker who wished to completely avoid detection, using a single account?
We consider a thought experiment where an attacker is given information for whether or not each query to the detection scheme will be detected as part of an attack sequence and knows $k$. 
Normally, the attacker will not know this information, as the encoder is not made public in our scheme.
How should an adversary proceed to optimally generate an adversarial example in this setting?
Recall that we proposed two instantiations of our defense: a time-bounded history of prior queries, or a history storing a fixed number of queries (e.g., in a circular buffer).
We analyze both schemes below.

\paragraph{Case 1: time-bounded buffer.}
Suppose we store each query until $t$ hours have passed.
Then, an attack that triggers $d$ detections would take an attacker $dt$ hours to execute if the attacker wishes to avoid being detected.
The most effective attack against our scheme that we have found (NES with auto-encoder query blinding) triggers 97 detections, so if we store queries for 100 hours, generating a single adversarial example without being detected would take over a year.

In practice, the defender can set the buffer duration according to their resource capacity. For example, Google's Cloud Vision API currently has a rate limit of 1800 queries per minute per user, so a user who continuously makes the maximum number of queries could make 10.8M queries per 100 hours.
Our similarity encoder maps each input to a $256$-dimensional output,
so storing these $10.8$M vectors (with $16$-bit floating point values) would require $5.5$GB of storage.
Google's cloud storage is currently priced at \$0.026/GB/month per user, so the storage cost to the provider of the query buffer would be at most \$0.14/month per user, and far less for most users (as we expect few users continuously max out the rate limits).


\paragraph{Case 2: query-bounded buffer.}
Suppose we always store the last $N$ queries made by each user.
Then an attacker who knows which queries will trigger a detection could avoid detection by flushing the buffer (by making $N$ extra random queries) just before being detected.
With this strategy, an attack that triggers $d$ detections would take an attacker about $dN$ total queries to execute if the attacker wishes to avoid being detected.

For example, Google's Cloud Vision API currently costs \$1.50 USD per $1000$ queries.
If the buffer stores $N = 10^4$ examples, our most effective attack would require
about $97 \times 10^4$ queries to execute without being detected, which would cost about \$1500 USD, a sizeable amount of money for an adversary to pay for generating a single adversarial image.
For comparison, without our defense, the best attack we tried would cost about \$20 USD.
The cost to the provider of storing a buffer of size $N=10^4$ would be about \$0.0007/month per user, at Google cloud storage's current prices.

\section{Zero Query Defense} \label{section:zero-query-defense}

\begin{table*}
\begin{tabular}{llrrrl}
\toprule
                  &   Attack              & Attack Success Rate      & Num. Queries & Detections \\
                  \midrule
\multirow{5}{*}{} & NES (best)          & 17\%           & 22,510$\pm$17,000       & 442      \\
                  & Boundary (best)            & 95\%    & 19,928$\pm$24,300            &     391         \\
                  & Hybrid Surrogate (untargeted)  &  5\%     & 8,192+/-0                   & 82 \\
                  \bottomrule
\end{tabular}
\caption{The effectiveness of our defense with an EAT-defended model. Our defense still detects query-based attacks, and EAT makes the defended model robust to zero-query attacks. Each attack was allowed to make up to 200,000 queries, and was run on the same set of randomly selected 100 images and target classes. The number of queries and detections are an average over the successful attacks.}
  \label{table:eat-results}
\end{table*}

Our scheme detects query-based attacks, but cannot detect zero-query attacks.
We propose that our scheme be combined with an existing defense against black-box zero-query attacks.
Currently,ensemble adversarial training (EAT) is one of the most effective defenses against black-box zero-query attacks \cite{EAT}.
EAT generates white-box adversarial examples with distortion $\epsilon$ on an ensemble of static models with different architectures and weights, and trains the defended model on these examples.
This procedure has been demonstrated to make the defended model robust against adversarial examples transferred from a holdout surrogate model, while resulting in only a modest decrease in the defended model's clean accuracy (on non-adversarial examples). Accordingly, we train a ResNet50v1 classifier for CIFAR-10 using EAT, with $\epsilon = 0.05$, to construct a model robust to zero-query attacks of $\epsilon = 0.05$.  Further training details can be found in the appendix.

To evaluate the EAT-defended model, ResNet50v1-EAT, against zero-query transfer attacks, we generated adversarial examples (over the CIFAR-10 test set) on a holdout ResNet74v2 model, and then saw if those examples transferred (i.e., were also adversarial for the ResNet50v1-EAT).
To give the attacker every advantage, we trained the ResNet74v2 surrogate model on the same CIFAR-10 training set, with the same training parameters. We used FGSM \cite{AdvExamples} and a clipped modification of the Carlini-Wagner (CW) $\ell_2$ attack \cite{Carlini} to generate adversarial examples (with $\epsilon = 0.05$ and $\kappa = 100$ for the CW attack).
Table~\ref{table:eat-transfer} shows the success rate of attacks against the defended ResNet50v1-EAT.

\begin{table}
\begin{tabular}{lrrrr}
\toprule
     Model           &    Clean      & FGSM(u)   & CW(u) & CW(t)   \\
                \midrule
ResNet50v1      & 7.8\%           & 73.8\%   & 59.1\%                         & 18.8\% \\
ResNet50v1-EAT  & 14.6\%          & 14.4\%   & 16.4\%                         & 1.0\%          \\
\bottomrule
\end{tabular}
\caption{Robustness against transfer attacks of an unprotected ResNet compared to an EAT-defended ResNet.  The second column shows the error rate on clean examples; the subsequent columns show the success rate of transfer attacks using untargeted FGSM, and untargeted and targeted variants of the attack from Carlini and Wagner \cite{Carlini}.}
 \label{table:eat-transfer}
\end{table}

Compared to the undefended model, the EAT-defended model is noticeably more robust to all three methods of transfer attack. The EAT-defended model does incur a noticeable decrease in clean accuracy, but in exchange we obtain substantial robustness against transfer attacks for a fairly large value of $\epsilon = 0.05$ (for reference, \cite{EAT} used $\epsilon = 0.06$ for defending models trained for much higher dimensional, 256x256x3 ImageNet images). In practice, the defender may tune $\epsilon$ according to their demands between accuracy and robustness. 

Our defense remains effective an EAT-defended model. We reran the best query-based attack variants on this EAT-defended model; results are shown in Table \ref{table:eat-results}. Our scheme is still able to detect query-based attacks frequently, and it appears that the EAT model may even reduce the success rate of query-based attacks.



\section{Related Work}

To our knowledge, our scheme is the first to use the history of queries in order to detect query-based black-box attacks for creating adversarial examples.
The most closely related work is PRADA \cite{PRADA}, which detects black-box model extraction attacks using the history of queries.
They examine the $\ell_2$ distance between images and raising an alarm if the distribution of these distances is not Gaussian.
Because they use $\ell_2$ distance on images, their scheme is not robust to query blinding, and because they examine only the distribution of distances, their scheme is not robust to insertion of dummy queries to make the distribution Gaussian \cite[\S V.C]{PRADA}.
They do not consider how to detect creation of adversarial examples.

Much previous work has explored stateless detection of whether an individual query is adversarial, usually by checking if the query is out of the distribution of normal/benign data \cite{metzen2017detecting, feinman2017detecting, grosse2017detection}. However, effective detection under this stateless threat model has proven difficult \cite{carlini2017adversarial}.

Other work has been done to defend against white-box attacks, such as adversarial training \cite{madry2017towards}. Such defenses are complementary to our defense: we can apply our detection strategy on top of any model. In our paper we study our defense on top of a non-robust model for simplicity and to accurately measure the value of this type of defense.
Recent work on robust similarity~\cite{elpips} could also be useful for improving our scheme.

There are other query-based black-box attacks, but they often follow either a similar gradient-estimation approach as NES, or a boundary-following approach similar to the boundary attack. For example, SPSA \cite{SPSA}, another gradient-estimation attack, estimates the gradient with Bernoulli instead of Gaussian directions. Boundary attack variants, like Boundary++ \cite{Boundary++}, RED \cite{RED}, and qFool \cite{qFool}, are more query efficient but still follow the same core approach of querying along the boundary.

Transfer attacks are a common approach in the zero-query setting \cite{papernot2016transfer}. We explore combining our defense with ensemble adversarial training \cite{EAT}, currently one of the most effective defenses against zero-query transfer attacks, but the recent Sitatapatra defense may also be effective \cite{anderson2019transfer}.

The approach for query blinding takes inspiration from previous work in signature blinding \cite{chaum1983blind} and mimicry attacks \cite{wagner2002mimicry}.

\section{Limitations and Future Work}


Our defense assumes that the model can be kept secret.
If the attacker can learn the model weights (e.g., through a model extraction attack \cite{ModelExtraction}) then our proposed defense again is not effective.
It would be interesting to explore stateful methods to detect model extraction attacks \cite{PRADA}.

We evaluated our defense on an image recognition task.
It would be interesting to explore whether similar ideas can be applied to other application domains as well, by constructing a suitable similarity encoder for that domain.

Our proposed defense only prevents attacks which attempt to generate an adversarial example near a specific image $x$.
If the adversary were content with finding \emph{any} image that is misclassified, then an adversary could simply generate random candidate inputs and wait until a test error randomly occurred \cite{ford2019adversarial}.
For example, because our CIFAR-10 classifier reaches $92\%$ accuracy an adversary who wanted to identify any error would be expected to succeed after trying just 13 images.
This attack is not specific to our defense, and would work equally well on any other defense to adversarial examples.

The specific auto-encoder attack transformation we develop is explicitly designed to target the defense we have constructed, and likely future defenses could detect this specific attack strategy.
However, we believe the general query-blinding strategy is an interesting research direction to pursue to develop stronger attacks.

Our proposed defense was only evaluated in the ``hard-label'' setting where the adversary receives only the classification label, without confidence scores.
Designing a defense in the ``soft-label'' black-box setting, where the model does return confidence scores, is an interesting direction for future work. In Appendix E we give a proposed attack which which demonstrates the potential difficulties in solving this case.

It may be possible to train a stronger defense that achieves higher robustness against query-blinding attacks through adversarial training.
We have a similarity encoder that is designed to detect similar images,
and an auto-encoder that is designed to produce images that fool the similarity
detector.
Similar to GANs, it is possible that jointly and adversarially training them might provide make the similarity encoder more robust.

\section{Conclusion}
Defenses against white-box adversarial examples have thus far proven elusive;
thus, we advocate for increased study into black-box defenses against adversarial examples.
In the black-box setting, the academic community has thus far studied only stateless defenses; we argue that stateful defenses give the defender a new advantage and deserve attention. 
Towards this end, we propose a simple scheme that detects the process of adversarial example generation.
By combining our proposed approach with existing defenses that
prevent transferability attacks, we construct the first unified defense that might offer black-box robustness.

\section*{Acknowledgements}
This work was supported by generous gifts from Google and Futurewei and
by the Hewlett Foundation through the Center for Long-term Cybersecurity.

\bibliographystyle{plain}
\bibliography{works}
\nocite{*}

\appendix
\section{CNN Architecture}
\begin{table}[!h]
\begin{center}
\begin{tabular}{llrp{2.5cm}}
\toprule
Layer 	&  	 	Filters 	& 	Size  	&  	Details\\
\midrule
Conv  	&  	32 		& 	$3 \times 3$	&  ReLU \\
Conv  & 	32 		&	$3 \times 3 $ 	& ReLU  \\
Max Pool  &   		& 	$2 \times 2$	& stride = 2 \\
Dropout  &   		& 		& $p = 0.25$\\
Conv  	&  	64 		& 	$3 \times 3$	&  ReLU \\
Conv  & 	64 		&	$3 \times 3 $ 	& ReLU  \\
Max Pool  &   		& 	$2 \times 2$	& stride = 2 \\
Dropout  &   		& 		& $p = 0.25$\\
Dense &  & 512 & ReLU\\
Dropout & &  & $p = 0.5 $\\
Dense &  & 256 & \\
\bottomrule

\end{tabular}
\caption{\textbf{Architecture for three layer CNN used for similarity encoder (section 3.4) and hybrid attack (section 6.3).}}
\label{tab:ArchDescription}
\end{center}
\end{table}

\section{Query blinding}
\label{sec:transform-params}

\subsection{Transformation Parameters}
Parameters for the low-distortion transformations (expected $\ell_2 = 2.32$), their parameters ($r$) are listed in Table~\ref{tab:transform-params-low}, and those for high-distortion transformations (expected $\ell_2 = 5.10$) are in Table~\ref{tab:transform-params-high}.

\begin{table}
\begin{center}
\begin{tabular}{llrp{2.5cm}}
\toprule
Transform 	&  	 	$r$ \\
\midrule
Uniform Noise  	&  	0.064 	 \\
Translate  	&  	0.45 	 \\
Rotate  	&  0.018 	 \\
Pixel-wise Scale  	&  	0.17 	 \\
Crop and Resize  	&  	0.04 	 \\
Brightness  	&  	0.09 	 \\
Contrast  	&  	0.55	 \\
Gaussian Noise  	&  	0.095 	 \\
\bottomrule

\end{tabular}
\caption{\textbf{Transformation parameters (low distortion)}}
\label{tab:transform-params-low}
\end{center}
\end{table}

\begin{table}
\begin{center}
\begin{tabular}{l|lrp{2.5cm}}
\toprule
Transform 	&  	 	$r$ \\
\midrule
Pixel-wise Scale  	&  	0.36 	 \\
Brightness  	&  	0.204 	 \\
Contrast  	&  	0.79	 \\
\bottomrule

\end{tabular}
\caption{\textbf{Transformation parameters (high distortion)}}
\label{tab:transform-params-high}
\end{center}
\end{table}

\subsection{NES Parameters}
We also explored reducing $s$, the number of samples per confidence score estimation, to $s = 2$ versus the original value $s = 50$, as a lower $s$ should result in less groups of $k = 50$ neighboring points being significantly near each other and accordingly less detections. This comes at a cost of potentially less accurate estimations of the confidence scores, so the adjustment of $s$ is experimented with in combination with query blinding, whereby different transformations (some that may be more accurate for estimating the scores than others) are used to sample points for estimating the scores. Additionally, a higher value of $s$ increases the chance that the attack will succeed, at the cost of increasing the number of detections due to the corresponding higher number of queries. The comparison of $s = 2$ versus $s = 50$, allowing up to $200,000$ or $5,000,000$ queries respectively, over different image transformations is shown in Figure \ref{table:NEStransformResultsExtended}. Each attack was run on the same set of randomly selected 100 images and target classes, and each metric is the average over the successful attacks.

\begin{figure*}
	\begin{tabular}{lllllll}
                  &  \emph{$s=2$:}               & Attack Success Rate  & Num. of Queries  &$\ell_2$ Detections & Similarity Encoding Detections \\
                  \toprule
\multirow{8}{*}{} & Uniform Noise   & 1\%                 &  15,695          & 308           & 308                               \\
                  & Translate       & 4\%                 &   6,714       & 131           & 131                               \\
                  & Rotate          & 7\%                 &  10,175      & 199           & 198                               \\
                  & Scale           & 27\%                & 12,567       & 246           & 246                                \\
                  & Crop and Resize &  7\%               & 6,132        & 119           & 119                                \\
                  & Brightness      & 55\%               & 13,503       & 263          & 264                                \\
                  & Contrast        & 23\%                & 11,197       & 219           & 219                                \\
                  & Gaussian Noise  & 0\%                 & n/a           & n/a           & n/a \\
                  \bottomrule
		\end{tabular}
		
		\begin{tabular}{lllllll}
                  &  \emph{$s=50$:}               & Attack Success Rate  & Num. of Queries  &$\ell_2$ Detections & Similarity Encoding Detections \\
                  \toprule
\multirow{8}{*}{} & Uniform Noise   & 15\%                &  453,651          & 8,895           & 8,895                               \\
                  & Translate       & 28\%                &   293,586       & 5,755           & 5,756                               \\
                  & Rotate          & 50\%                 &  295,022      & 5,782           & 5,784                               \\
                  & Scale           & 83\%               & 337,346       & 6,607           & 6,614                                \\
                  & Crop and Resize &  43\%             & 249,717        & 4,895           & 4,895                                \\
                  & Brightness      & 45\%              & 277,864       & 5,437          & 5,447                                \\
                  & Contrast        & 23\%                & 253,325       & 4,966           & 4,966                                \\
                  & Gaussian Noise  & 0\%                & n/a           & n/a           & n/a                                    
		\end{tabular}
  \caption{\textbf{Attack success and detection rate between $s = 2$ and $s = 50$.}}
  \label{table:NEStransformResultsExtended}
\end{figure*}


\section{Ensemble Adversarial Training}
We pre-trained a ResNet50v1 on CIFAR-10 (accuracy 92.2\%).
Then we train it on adversarial examples generated on an ensemble of trained ResNets with different architectures: ResNet44v1, ResNet56v2, and ResNet74v1. Adversarial examples were generated using FGSM \cite{AdvExamples} with $\epsilon = 0.05$. The network was trained for 100 epochs, where the adversarial examples for each epoch were generated from a randomly selected model from the ensemble and the defended model, and one adversarial example was generated per image in the CIFAR-10 training set. 

\newpage
\section{A Difficult Soft-Label Case}
\label{sec:softlabel}
Our paper focuses on the ``hard-label'' threat model where the adversary only
is only given the label with highest confidence score.
In contrast, the ``soft-label'' threat model gives the adversary full access to
the output probabilities of the model (the full output of the softmax layer). 
%
%
We present one possible attack that demonstrates the increased capabilities for
an attacker when given access to the probability distribution.
In particular, this attack shows that our hard-label defense scheme does not scale to
the soft-label setting.

To obtain the classification of a sample $x$ we define the blinding and revealing functions as follows:
\begin{itemize}
    \item To blind $x$, select a random unit-vector direction $r$ and let
    \[ b(x) = \{x + d \cdot r, x + (d+\epsilon) \cdot r \} \]
    where the distance $d$ is a hyperparameter of the attack.
    \item Recover $f(x)$ by letting
    \[ r(y_0, y_1) = y_0 + {d \over \epsilon} \cdot (y_0 - y_1). \]
\end{itemize}
This attack works due to the fact that neural networks, despite being non-linear functions globally, often locally behave as if they were linear functions.

To increase the accuracy of the estimate of $f(x)$ the above procedure can be generalized by selecting multiple random directions $r_i$ and averaging the results.
We experimentally verified that this type of attack approach is effective and allows the soft-label variant of NES \cite{NES} to succeed against our query-based detector.

\end{document}